\documentclass[aps,preprint,tightenlines,showpacs,nofootinbib]{revtex4}
\usepackage{epsfig}
\usepackage{multirow}
\usepackage{amsmath,stackrel}
\begin{document}
\title{$N\bar D $ system: A challenge for $\bar {\rm P}$ANDA}
\author{T.~F.~Caram\'es}
\affiliation{Departamento de F{\'\i}sica Fundamental,
Universidad de Salamanca, 37008 Salamanca, Spain}
\author{A.~Valcarce}
\affiliation{Departamento de F{\'\i}sica Fundamental,
Universidad de Salamanca, 37008 Salamanca, Spain}
\date{\emph{Version of }\today}

\begin{abstract}
We study the $N \bar D$ system by means of a chiral constituent quark model. 
This model, tuned in the description of the baryon and meson spectra 
as well as the $NN$ interaction, provides parameter-free predictions for 
the charm $-1$ two-hadron systems. The presence of a heavy antiquark makes the 
interaction rather simple. We have found sharp quark-Pauli effects in some 
particular channels, generating 
repulsion, due to lacking degrees of freedom to accommodate the light quarks.
Our results point to the existence of an attractive state, the 
$\Delta \bar D^*$ with $(T,S)=(1,5/2)$, presenting a resonance close to
threshold. It would show up 
in the scattering of $\bar D$ mesons on nucleons as a D wave state 
with quantum numbers $(T)J^P=(1)5/2^-$. This resonance resembles
our findings in the $\Delta\Delta$ system, that offered a plausible explanation to the cross section
of double-pionic fusion reactions through the so-called
{\it ABC} effect. The study of the interaction of $D$ mesons with nucleons is a goal of the
$\bar {\rm P}$ANDA Collaboration at the European facility FAIR and, thus, the present
theoretical study is a challenge to be tested at the future experiments.
\end{abstract}

\pacs{14.40.Lb,12.39.Pn,12.40.-y}
\maketitle

\section{Introduction}
\label{secI}

The study of the interaction between charmed mesons and nucleons has become an interesting
subject in several contexts~\cite{Pan09}. It is particularly interesting for the study
of chiral symmetry restoration in a hot and/or dense medium~\cite{Kog83}. It will also help in the
understanding of the suppression of the $J/\Psi$ production in heavy ion collisions~\cite{Mat86}. Besides,
it may shed light on the possible existence of exotic nuclei with heavy flavors~\cite{Dov77}. Experimentally, 
it will become possible to analyze the interaction of charmed mesons with nucleons inside 
nuclear matter with the operation of the FAIR facility at the GSI laboratory in Germany~\cite{Pan09}. 
There are proposals for experiments by the $\bar {\rm P}$ANDA Collaboration to produce $D$ mesons by 
annihilating antiprotons on the deuteron. This could be achieved with an antiproton beam, by tuning 
the antiproton energy to one of the higher-mass 
charmonium states that decays into open charm mesons. The $\bar D$ mesons then have a chance to 
interact with the nucleons inside the target material. Since the $D$ mesons are produced in pairs 
in the antiproton-nucleon annihilation process, the appearance of one of those $D$ mesons can be 
used as tag to look for such reactions. These experimental ideas may become plausible based on 
recent estimations of the cross section for the production of $D\bar D$ pairs in proton-antiproton
collisions~\cite{Kho12}. Thus, a good knowledge of the interaction of charmed mesons 
with ordinary hadrons, like nucleons, is a prerequisite. 

Before one can infer in a sensitive way changes of the interaction in the medium, a reasonable 
understanding of the interaction in free space is required. However, here one has to manage
with an important difficulty, namely the complete lack of experimental data at low energies for the
free-space interaction. Thus, the generalization of models describing the two-hadron interaction
in the light flavor sector could offer insight about the unknown interaction of hadrons with
heavy flavors. This is the main purpose of this work, to make use of a chiral constituent quark
model describing the $NN$ interaction~\cite{Val05} as well as the meson spectrum in all 
flavor sectors~\cite{Vij05} to obtain parameter-free predictions that may be testable in 
the future experiments of the $\bar {\rm P}$ANDA Collaboration. Such a project was already
undertaken for the interaction between charmed mesons with reasonable predictions~\cite{Car09},
what encourages us in the present challenge.

The paper is organized as follows. In Sec.~\ref{secII} we present a description
of the quark-model wave function for the baryon-meson system, centering our attention
in its short-range behavior looking for quark-Pauli effects. In Sec.~\ref{secIII} we
briefly revise the interacting potential. Section~\ref{secIV} deals with
the solution of the two-body problem by means of the Fredholm determinant.
In Sec.~\ref{secV} we present our results. We will discuss in detail the baryon-meson 
interactions emphasizing those aspects that may produce different results from
purely hadronic theories. We will analyze the character of the interaction in the 
different isospin-spin channels, looking for the attractive ones that may lodge 
resonances to be measured at $\bar {\rm P}$ANDA.
We will also compare with existing results in the literature.
Finally, in Sec.~\ref{secVI} we summarize our main conclusions.

\section{The baryon-meson wave function}
\label{secII}

In order to describe the baryon-meson system we shall use a constituent quark
cluster model, i.e., hadrons are described as clusters of
quarks and antiquarks. Assuming a two-center shell model the wave function of
an arbitrary baryon-meson system, a baryon $B_i$ and a meson $M_j$, can be 
written as:
\begin{equation}
\Psi_{B_i M_j}^{LST}({\vec R}) =  {\cal A}
\left[ B_i \left( 123;{-{\frac{{\vec R} 
}{2}}} \right) M_j \left( 4\bar{5}; {+\frac{{\vec R} }{2}} \right) \right]^{LST}
\, ,  \label{Gor}
\end{equation}
where ${\cal A}$ is the antisymmetrization operator accounting for 
the possible existence of identical quarks inside the hadrons. In the case we are
interested in, baryon-meson systems made of $N$ or $\Delta$ baryons and 
$\bar D$ or $\bar D^*$ mesons, the antisymmetrization operator is given by
\begin{equation}
{\cal A} = \left ( 1-\sum_{i=1}^3 P^{LST}_{ij} \right ) \, ,
\end{equation}

\noindent where $P^{LST}_{ij}$ exchanges a pair of identical quarks $i$ and $j$, and 
$j$ stands for the light quark of the charmed meson.
If we assume gaussian $0s$ wave functions for the quarks inside the hadrons,
the normalization of the baryon-meson wave function $\Psi_{B_i M_j}^{LST}({\vec R})$
of Eq.~(\ref{Gor}) can be expressed as,

\begin{equation}
{\cal N}_{B_iM_j}^{LST}(R)= {\cal N}^{L}_{\rm di}(R) - C(S,T) {\cal N}%
^{L}_{\rm ex}(R) \, .
\label{Norm}
\end{equation}

\noindent 
${\cal N}_{\rm di}^{L}(R)$ and ${\cal N}_{\rm ex}^{L}(R)$ stand for
the direct and exchange radial normalizations, respectively,
whose explicit expressions are
\begin{eqnarray}
 {\cal{N}}_{\rm di}^{L} (R) &=& 4 \pi \exp \left\lbrace {-\frac{R^2}{8} 
\left( \frac{4}{b^2} + \frac{1}{b_c^2}\right)}\right\rbrace  i_{L+1/2} 
\left[ \frac{R^2}{8} \left( \frac{4}{b^2} + \frac{1}{b_c^2} \right)\right] \label{nex12}\,, \\
{\cal{N}}_{\rm ex}^{L} (R) &=& 4 \pi \exp \left\lbrace {-\frac{R^2}{8} 
\left( \frac{4}{b^2} + \frac{1}{b_c^2}\right)}\right\rbrace  i_{L+1/2} 
\left[ \frac{R^2}{8 b_c^2} \right] \,, \nonumber
\label{Norm2}
\end{eqnarray}
where, for the sake of generality, we have assumed different gaussian parameters
for the wave function of the light quarks ($b$) and the heavy quark ($b_c$).
In the limit where the two hadrons overlap ($R \to 0$), the Pauli principle
may impose antisymmetry requirements not present in a hadronic description.
Such effects, if any, will be prominent for $L=0$. Using the asymptotic form
of the Bessel functions, $i_{L+1/2}$, we obtain,
\begin{eqnarray}
{\cal{N}}_{\rm di}^{L=0} &\stackrel[R\to 0]{}{\hbox to 20pt{\rightarrowfill}}& 4 \pi \left\lbrace {1 -\frac{R^2}{8} 
\left( \frac{4}{b^2} + \frac{1}{b_c^2}\right)}\right\rbrace  
\left[1 + \frac{1}{6} \left( \frac{R^2}{8} \left( \frac{4}{b^2} + \frac{1}{b_c^2} \right) \right)^2 + ...\right] \nonumber\,, \\
{\cal{N}}_{\rm ex}^{L=0} &\stackrel[R\to 0]{}{\hbox to 20pt{\rightarrowfill}}& 4 \pi \left\lbrace {1 -\frac{R^2}{8} 
\left( \frac{4}{b^2} + \frac{1}{b_c^2}\right)}\right\rbrace  
\left[1 + \frac{1}{6} \left( \frac{R^2}{8 b_c^2}\right)^2 + ...\right]  \,. \label{npro}
\end{eqnarray}
Finally, the S wave normalization kernel, Eq.~(\ref{Norm}), can be written in the overlapping region
as
\begin{equation}
{\cal N}_{B_iM_j}^{L=0ST} \stackrel[R\to 0]{}{\hbox to 20pt{\rightarrowfill}} 4\pi \left\lbrace {1 -\frac{R^2}{8} 
\left( \frac{4}{b^2} + \frac{1}{b_c^2}\right)}\right\rbrace  
\left\lbrace \left[ 1 - C(S,T) \right] +  \frac{1}{6} \left( \frac{R^2}{8 b_c^2}\right)^2  
\left[ \gamma^2 - C(S,T) \right] + ... \right\rbrace \,,
\end{equation}
where $\gamma=1+\frac{4b_c^2}{b^2}$.
Thus, the closer the value of $C(S,T)$ to 1 the larger the suppression of the normalization
of the wave function at short distances, generating Pauli repulsion. In particular,
if $C(S,T)=1$ the norm goes to zero for $R \to 0$, what is called Pauli blocking~\cite{Val97}.
$C(S,T)$ is a coefficient depending on the total spin ($S$) and isospin ($T$) of the 
$B_iM_j$ two-hadron system and is given by
\begin{table}[b]
\caption{$C(S,T)$ spin-isospin coefficients defined in Eq.~(\ref{Cst}).}
\label{tab1}
\begin{tabular}{|c|c|ccc|}
\hline
&$B_i M_j$ & $T=0$  &  $T=1$ & $T=2$ \\
\hline\hline
\multirow{3}{*}{$S=1/2$} & $N \bar D$ &  $0$ &  $1/3$ & $-$ \\
& $N \bar D^* $ &  $2/3$ &   $-1/9$ & $-$ \\
& $\Delta \bar D^* $ & $-$ &  $1/9$ & $-1/3$ \\ \hline
\multirow{3}{*}{$S=3/2$} & $N \bar D^*$ & $-1/3$ & $5/9$ & $-$ \\
& $\Delta \bar D$ & $-$ & $-1/6$ & $1/2$ \\
& $\Delta \bar D^* $ & $-$ & $-1/18$ & $1/6$\\ \hline
$S=5/2$ & $\Delta \bar D^* $ & $-$ & $-1/3$ & $1$ \\ \hline
\end{tabular}
\end{table}
\begin{figure}[t]
\vspace*{-2cm}
\hspace*{-1cm}\mbox{\epsfxsize=160mm\epsffile{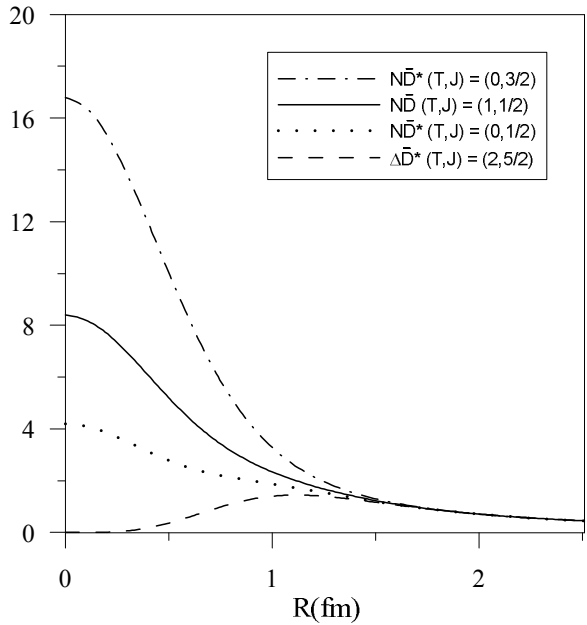}}
\vspace*{-13.5cm}
\caption{Norm kernel defined in Eq.~(\protect{\ref{Norm}}) for $L=0$ and four different channels.}
\label{fig1}
\end{figure}
\begin{eqnarray}
C(S,T) & = & 3 \, \left( \frac{2S_1+1}{4}\right)  \nonumber \\
&&\sum_{\chi_i=\eta_i=0}^{1} \left<
\left( \chi_1, {1 \over 2} \right), {S_1} ;
\left( {1 \over 2}, {1 \over 2} \right), {S_2} ;
S, M_S \right| P_{k\ell}^S \left|
\left( \chi_2, {1 \over 2} \right), {S_1} ;
\left( {1 \over 2}, {1 \over 2}\right), {S_2} ;
S, M_S \right> \nonumber \\
& & \left<
\left( \eta_1, {1 \over 2} \right), {T_1} ;
\left( 0, {1 \over 2} \right), {1 \over 2} ;
T, M_T \right| P_{k\ell}^T \left|
\left( \eta_2, {1 \over 2} \right), {T_1} ;
\left( 0, {1 \over 2} \right), {1 \over 2} ;
T, M_T \right> \, ,
\label{Cst}
\end{eqnarray}
where the subindices $k$ and $\ell$ in the exchange operator $P^{ST}$ denote
a quark of the baryon $B_i$ and a quark of the meson $M_j$, respectively.
$\chi_i$ ($\eta_i$) stands for the
coupled spin (isospin) of two quarks inside the baryon, 
$(S_1,T_1)$ are the spin and isospin of the baryon $N$ or $\Delta$,
and $S_2$ is the spin of the meson $\bar D$ or $\bar D^*$.

The values of $C(S,T)$ are given in Table~\ref{tab1}. Similarly to
Pauli blocked channels, corresponding to $C(S,T)$=1, we will call 
Pauli suppressed channels those where $C(S,T)$ is close to one. 
We can see that there is one system
showing Pauli blocking, this is the $\Delta \bar D^*$ with $(T,S)=(2,5/2)$.
This is easily understood due to lacking degrees of freedom to
accommodate the light quarks present on this configuration. 
The interaction in this channel will be strongly repulsive~\cite{Val97}, as we will 
discuss in Sec.~\ref{secV}. The channel $N \bar D^*$ with $(T,J)=(0,1/2)$ where $C(S,T)=2/3$
is Pauli suppressed, the norm kernel gets rather small at short distances giving rise
to Pauli repulsion at short distances as we will see in Sec.~\ref{secV}. 
We show in Fig.~\ref{fig1} the normalization kernel given 
by Eq.~(\ref{Norm}) for $L=0$ and four different channels:
$\Delta \bar D^*$ with $(T,J)=(2,5/2)$,
$N \bar D^*$ with $(T,J)=(0,1/2)$,
$N \bar D$ with $(T,J)=(1,1/2)$, and
$N \bar D^*$ with $(T,J)=(0,3/2)$. In the first three cases $C(S,T)$
is positive, being exactly one in the first case and becoming smaller
in the others, what makes the norm kernel to augment. In the last case
$C(S,T)$ is negative, showing a large norm kernel at short distances 
and therefore one does not expect any Pauli effect at all.

\section{The two-body interactions}
\label{secIII}

The two-body interactions involved in the study of the baryon-meson
system are obtained from the chiral constituent quark model~\cite{Val05}. 
This model was proposed in the early 90's in an attempt to
obtain a simultaneous description of the nucleon-nucleon
interaction and the baryon spectra. It was later on generalized to all 
flavor sectors ~\cite{Vij05}. 
In this model baryons are described as clusters of three interacting 
massive (constituent) quarks, the mass coming from the spontaneous breaking 
of the original $SU(2)_{L}\otimes SU(2)_{R}$ 
chiral symmetry of the QCD Lagrangian.
QCD perturbative effects are taken into account
through the one-gluon-exchange (OGE) potential~\cite{Ruj75}.
It reads,  
\begin{equation}
V_{\rm OGE}({\vec{r}}_{ij})=
        {\frac{\alpha_s}{4}}\,{\vec{\lambda}}_{i}^{\rm
c} \cdot {\vec{\lambda}}_{j}^{\rm c}
        \Biggl \lbrace{ \frac{1}{r_{ij}}}
        - \dfrac{1} {4} \left(
{\frac{1}{{2\,m_{i}^{2}}}}\, +
{\frac{1}{{2\,m_{j}^{2}}}}\,
        + {\frac{2 \vec \sigma_i \cdot \vec
\sigma_j}{3 m_i m_j}} \right)\,\,
          {\frac{{e^{-r_{ij}/r_{0}}}}
{{r_{0}^{2}\,\,r_{ij}}}}
        - \dfrac{3 S_{ij}}{4 m_q^2 r_{ij}^3}
        \Biggr \rbrace\,\, ,
\end{equation}
where $\lambda^{c}$ are the $SU(3)$ color matrices, 
$r_0=\hat r_0/\mu$ is a flavor-dependent regularization scaling with the 
reduced mass of the interacting pair, and $\alpha_s$ is the
scale-dependent strong coupling constant given by~\cite{Vij05},
\begin{equation}
\alpha_s(\mu)={\alpha_0\over{\rm{ln}\left[{({\mu^2+\mu^2_0})/
\gamma_0^2}\right]}},
\label{asf}
\end{equation}
where $\alpha_0=2.118$, 
$\mu_0=36.976$ MeV and $\gamma_0=0.113$ fm$^{-1}$. This equation 
gives rise to $\alpha_s\sim0.54$ for the light-quark sector
and $\alpha_s\sim0.43$ for $uc$ pairs.

Non-perturbative effects are due to the spontaneous breaking of the original 
chiral symmetry at some momentum scale. In this domain of momenta, light quarks 
interact through Goldstone boson exchange potentials,
\begin{equation}
V_{\chi}(\vec{r}_{ij})\, = \, V_{\rm OSE}(\vec{r}_{ij}) \, + \, V_{\rm OPE}(\vec{r}_{ij}) \, ,
\end{equation}
where
\begin{eqnarray}
V_{\rm OSE}(\vec{r}_{ij}) &=&
    -\dfrac{g^2_{\rm ch}}{{4 \pi}} \,
     \dfrac{\Lambda^2}{\Lambda^2 - m_{\sigma}^2}
     \, m_{\sigma} \, \left[ Y (m_{\sigma} \,
r_{ij})-
     \dfrac{\Lambda}{{m_{\sigma}}} \,
     Y (\Lambda \, r_{ij}) \right] \,, \nonumber \\
V_{\rm OPE}(\vec{r}_{ij})&=&
     \dfrac{ g_{\rm ch}^2}{4
\pi}\dfrac{m_{\pi}^2}{12 m_i m_j}
     \dfrac{\Lambda^2}{\Lambda^2 - m_{\pi}^2}
m_{\pi}
     \Biggr\{\left[ Y(m_{\pi} \,r_{ij})
     -\dfrac{\Lambda^3}{m_{\pi}^3} Y(\Lambda
\,r_{ij})\right]
     \vec{\sigma}_i \cdot \vec{\sigma}_j 
\nonumber \\
&&   \qquad\qquad +\left[H (m_{\pi} \,r_{ij})
     -\dfrac{\Lambda^3}{m_{\pi}^3} H(\Lambda
\,r_{ij}) \right] S_{ij}
     \Biggr\}  (\vec{\tau}_i \cdot \vec{\tau}_j)
\, .
\end{eqnarray}
$g^2_{\rm ch}/4\pi$ is the chiral coupling constant,
$Y(x)$ is the standard Yukawa function defined by $Y(x)=e^{-x}/x$,
$S_{ij} \, = \, 3 \, ({\vec \sigma}_i \cdot
{\hat r}_{ij}) ({\vec \sigma}_j \cdot  {\hat r}_{ij})
\, - \, {\vec \sigma}_i \cdot {\vec \sigma}_j$ is
the quark tensor operator, and $H(x)=(1+3/x+3/x^2)\,Y(x)$.

Finally, any model imitating QCD should incorporate
confinement. Being a basic term from the spectroscopic point of view
it is negligible for the hadron-hadron interaction. Lattice calculations 
suggest a screening effect on the potential when increasing the interquark 
distance~\cite{Bal01},
\begin{equation}
V_{\rm CON}(\vec{r}_{ij})=\{-a_{c}\,(1-e^{-\mu_c\,r_{ij}})\}(\vec{%
\lambda^c}_{i}\cdot \vec{ \lambda^c}_{j})\, .
\end{equation}
\begin{table}[t]
\caption{Quark model parameters.}
\label{tab2}
\begin{tabular}{cccc}
\hline
\hline
& $m_{u,d} ({\rm MeV})$ & 313 &  \\ 
& $m_c ({\rm MeV})$ & 1752 &  \\ 
& $b ({\rm fm})$ & 0.518 &  \\ 
& $b_c ({\rm fm})$ & 0.6 &  \\ 
& $\hat r_0$ (MeV fm) & 28.170 &  \\ 
& $a_c ({\rm MeV \, fm^{-1}})$ & 109.7 &  \\ 
& $g_{\rm ch}^2/(4\pi)$      & 0.54&  \\  
& $m_\sigma ({\rm fm^{-1}})$ & 3.42 &  \\ 
& $m_\pi ({\rm fm^{-1}})$ & 0.70 &  \\ 
& $\Lambda ({\rm fm^{-1}})$ & 4.2 & \\
& $a_c$ (MeV)             & 230&\\
& $\mu_c$ (fm$^{-1}$)&0.70&\\ 
\hline
\hline
\end{tabular}
\end{table}
Once perturbative (one-gluon exchange) and nonperturbative (confinement and
chiral symmetry breaking) aspects of QCD have been considered, one ends up with
a quark-quark interaction of the form 
\begin{equation} 
V_{q_iq_j}(\vec{r}_{ij})=\left\{ \begin{array}{ll} 
\left[ q_iq_j=nn \right] \Rightarrow V_{\rm CON}(\vec{r}_{ij})+V_{\rm OGE}(\vec{r}_{ij})+V_{\chi}(\vec{r}_{ij}) &  \\ 
\left[ q_iq_j=cn \right]  \Rightarrow V_{\rm CON}(\vec{r}_{ij})+V_{\rm OGE}(\vec{r}_{ij}) & 
\end{array} \right.\,,
\label{pot}
\end{equation}
where $n$ stands for the light quarks $u$ and $d$.
Notice that for the particular case of heavy quarks ($c$ or $b$) chiral symmetry is
explicitly broken and therefore boson exchanges do not contribute.
For the sake of completeness we compile the parameters of the model in Table \ref{tab2}.
The model guarantees a nice description of the baryon ($N$ and $\Delta$)~\cite{Valb05}
and the meson ($\bar D$ and $\bar D^*$) spectra~\cite{Vij05}. Let us also note that
the parameters of the model have been tuned in the meson and baryon spectra and
the $NN$ interaction, and therefore
the present calculation is free of parameters.

In order to derive the local $B_n M_m\to B_k M_l$ interaction from the
basic $qq$ interaction defined above, we use a Born-Oppenheimer
approximation. Explicitly, the potential is calculated as follows,
\begin{equation}
V_{B_n M_m (L \, S \, T) \rightarrow B_k M_l (L^{\prime}\, S^{\prime}\, T)} (R) =
\xi_{L \,S \, T}^{L^{\prime}\, S^{\prime}\, T} (R) \, - \, \xi_{L \,S \,
T}^{L^{\prime}\, S^{\prime}\, T} (\infty) \, ,  \label{Poten1}
\end{equation}
\noindent where
\begin{equation}
\xi_{L \, S \, T}^{L^{\prime}\, S^{\prime}\, T} (R) \, = \, {\frac{{\left
\langle \Psi_{B_k M_l}^{L^{\prime}\, S^{\prime}\, T} ({\vec R}) \mid
\sum_{i<j=1}^{5} V_{q_iq_j}({\vec r}_{ij}) \mid \Psi_{B_n M_m}^{L \, S \, T} ({\vec R%
}) \right \rangle} }{{\sqrt{\left \langle \Psi_{B_k M_l }^{L^{\prime}\,
S^{\prime}\, T} ({\vec R}) \mid \Psi_{B_k M_l }^{L^{\prime}\, S^{\prime}\, T} ({%
\vec R}) \right \rangle} \sqrt{\left \langle \Psi_{B_n M_m }^{L \, S \, T} ({\vec %
R}) \mid \Psi_{B_n M_m }^{L \, S \, T} ({\vec R}) \right \rangle}}}} \, .
\label{Poten2}
\end{equation}
In the last expression the quark coordinates are integrated out keeping $R$
fixed, the resulting interaction being a function of the baryon-meson relative 
distance. The wave function $\Psi_{B_n M_m}^{L \, S \, T}({\vec R})$ for the baryon-meson
system has been discussed in detail in Sec.~\ref{secII}.
\begin{figure}[t]
\mbox{\epsfxsize=100mm\epsffile{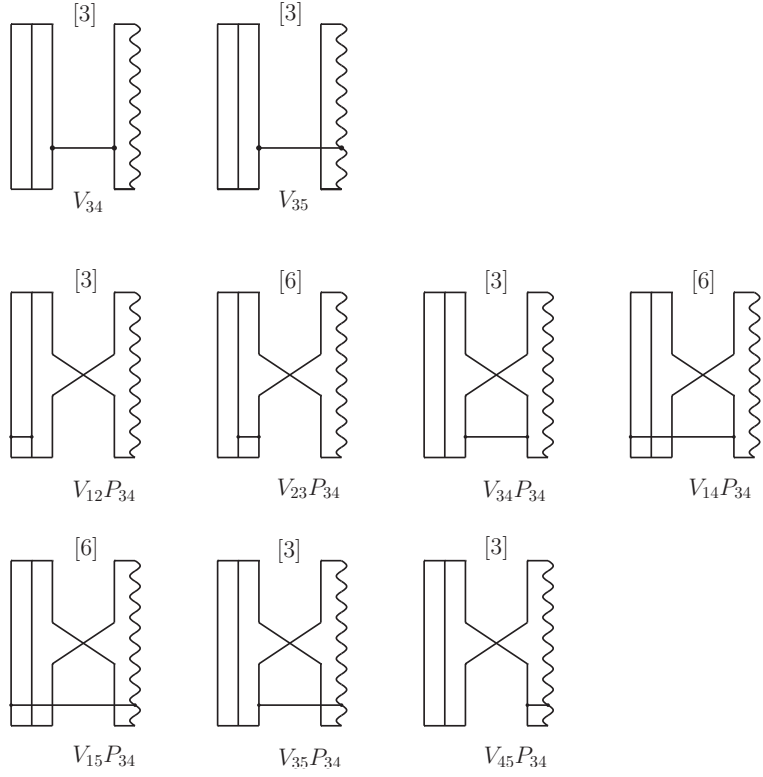}}
\caption{Different diagrams contributing to the
baryon-meson interaction. The vertical solid lines represent a light quark, $u$ or $d$,
while the wavy line represents the charm antiquark. The horizontal line denotes the
interacting quarks. The number between square brackets stands for
the number of diagrams topologically equivalent.}
\label{fig2}
\end{figure}
We show in Fig.~\ref{fig2} the different diagrams contributing to the baryon-meson interaction.
The contributions on the first line are pure hadronic interactions, while the diagrams in the
second and third lines contain quark-exchanges and therefore will not be present in a hadronic
description. In Sec.~\ref{secV} we will analyze in detail the different diagrams to make
clear the contributions arising from pure quark effects.

\section{Integral equations for the two-body systems}
\label{secIV}

To study the possible existence of exotic states made of a light baryon, $N$ or $\Delta$,
and a charmed meson, $\bar D$ or $\bar D^*$, we have solved the Lippmann-Schwinger equation 
for negative energies using the Fredholm determinant. This method permitted 
us to obtain robust predictions even for zero-energy bound states, and gave
information about attractive channels that may lodge a resonance~\cite{Car09}.
We consider a baryon-meson system $B_i M_j$ ($B_i=N$ or $\Delta$ and $M_j=\bar D$ 
or $\bar D^*$) in a relative $S$ state interacting through a potential $V$ that contains a
tensor force. Then, in general, there is a coupling to the 
$B_i M_j$ $D$ wave. Moreover, the baryon-meson system can couple to other 
baryon-meson states. We show in Table~\ref{tab3} the coupled channels 
in the isospin-spin $(T,J)$ basis. 
\begin{table}[b]
\caption{Interacting baryon-meson channels in the isospin-spin $(T,J$) basis.}
\label{tab3}
\begin{tabular}{cccc}
\hline \hline
& $T=0$  &  $T=1$ & $T=2$ \\
\hline
$J=1/2$ & $N \bar D  - N \bar D^* $         & $N \bar D - N \bar D^* - \Delta \bar D^* $         & $\Delta \bar D^*$ \\
$J=3/2$ & $N \bar D^* $                    & $N \bar D^* - \Delta \bar D - \Delta \bar D^* $  & $\Delta \bar D  - \Delta \bar D^*$ \\
$J=5/2$ &                & $\Delta \bar D^* $                               & $\Delta \bar D^* $ \\ \hline\hline
\end{tabular}
\end{table}
Thus, if we denote the different baryon-meson systems as channel $A_i$,
the Lippmann-Schwinger equation for the baryon-meson scattering becomes
\begin{eqnarray}
t_{\alpha\beta;TJ}^{\ell_\alpha s_\alpha, \ell_\beta s_\beta}(p_\alpha,p_\beta;E)& = & 
V_{\alpha\beta;TJ}^{\ell_\alpha s_\alpha, \ell_\beta s_\beta}(p_\alpha,p_\beta)+
\sum_{\gamma=A_1,A_2,\cdots}\sum_{\ell_\gamma=0,2} 
\int_0^\infty p_\gamma^2 dp_\gamma V_{\alpha\gamma;TJ}^{\ell_\alpha s_\alpha, \ell_\gamma s_\gamma}
(p_\alpha,p_\gamma) \nonumber \\
& \times& \, G_\gamma(E;p_\gamma)
t_{\gamma\beta;TJ}^{\ell_\gamma s_\gamma, \ell_\beta s_\beta}
(p_\gamma,p_\beta;E) \,\,\,\, , \, \alpha,\beta=A_1,A_2,\cdots \,\, ,
\label{eq0}
\end{eqnarray}
where $t$ is the two-body scattering amplitude, $T$, $J$, and $E$ are the
isospin, total angular momentum and energy of the system,
$\ell_{\alpha} s_{\alpha}$, $\ell_{\gamma} s_{\gamma}$, and
$\ell_{\beta} s_{\beta }$
are the initial, intermediate, and final orbital angular momentum and spin, respectively,
 and $p_\gamma$ is the relative momentum of the
two-body system $\gamma$. The propagators $G_\gamma(E;p_\gamma)$ are given by
\begin{equation}
G_\gamma(E;p_\gamma)=\frac{2 \mu_\gamma}{k^2_\gamma-p^2_\gamma + i \epsilon} \, ,
\end{equation}
with
\begin{equation}
E=\frac{k^2_\gamma}{2 \mu_\gamma} \, ,
\end{equation}
where $\mu_\gamma$ is the reduced mass of the two-body system $\gamma$.
For bound-state problems $E < 0$ so that the singularity of the propagator
is never touched and we can forget the $i\epsilon$ in the denominator.
If we make the change of variables
\begin{equation}
p_\gamma = d{1+x_\gamma \over 1-x_\gamma},
\label{eq2}
\end{equation}
where $d$ is a scale parameter, and the same for $p_\alpha$ and $p_\beta$, we can
write Eq.~(\ref{eq0}) as
\begin{eqnarray}
t_{\alpha\beta;TJ}^{\ell_\alpha s_\alpha, \ell_\beta s_\beta}(x_\alpha,x_\beta;E)& = & 
V_{\alpha\beta;TJ}^{\ell_\alpha s_\alpha, \ell_\beta s_\beta}(x_\alpha,x_\beta)+
\sum_{\gamma=A_1,A_2,\cdots}\sum_{\ell_\gamma=0,2} 
\int_{-1}^1 d^2\left(1+x_\gamma \over 1-x_\gamma \right)^2 \,\, {2d \over (1-x_\gamma)^2}
dx_\gamma \nonumber \\
&\times & V_{\alpha\gamma;TJ}^{\ell_\alpha s_\alpha, \ell_\gamma s_\gamma}
(x_\alpha,x_\gamma) \, G_\gamma(E;p_\gamma) \,
t_{\gamma\beta;TJ}^{\ell_\gamma s_\gamma, \ell_\beta s_\beta}
(x_\gamma,x_\beta;E) \, .
\label{eq3}
\end{eqnarray}
We solve this equation by replacing the integral from $-1$ to $1$ by a
Gauss-Legendre quadrature which results in the set of
linear equations
\begin{equation}
\sum_{\gamma=A_1,A_2,\cdots}\sum_{\ell_\gamma=0,2}\sum_{m=1}^N
M_{\alpha\gamma;TJ}^{n \ell_\alpha s_\alpha, m \ell_\gamma s_\gamma}(E) \, 
t_{\gamma\beta;TJ}^{\ell_\gamma s_\gamma, \ell_\beta s_\beta}(x_m,x_k;E) =  
V_{\alpha\beta;TJ}^{\ell_\alpha s_\alpha, \ell_\beta s_\beta}(x_n,x_k) \, ,
\label{eq4}
\end{equation}
with
\begin{eqnarray}
M_{\alpha\gamma;TJ}^{n \ell_\alpha s_\alpha, m \ell_\gamma s_\gamma}(E)
& = & \delta_{nm}\delta_{\ell_\alpha \ell_\gamma} \delta_{s_\alpha s_\gamma}
- w_m d^2\left(1+x_m \over 1-x_m\right)^2{2d \over (1-x_m)^2} \nonumber \\
& \times & V_{\alpha\gamma;TJ}^{\ell_\alpha s_\alpha, \ell_\gamma s_\gamma}(x_n,x_m) 
\, G_\gamma(E;{p_\gamma}_m),
\label{eq5}
\end{eqnarray}
and where $w_m$ and $x_m$ are the weights and abscissas of the Gauss-Legendre
quadrature while ${p_\gamma}_m$ is obtained by putting
$x_\gamma=x_m$ in Eq.~(\ref{eq2}).
If a bound state exists at an energy $E_B$, the determinant of the matrix
$M_{\alpha\gamma;TJ}^{n \ell_\alpha s_\alpha, m \ell_\gamma s_\gamma}(E_B)$ 
vanishes, i.e., $\left|M_{\alpha\gamma;TJ}(E_B)\right|=0$.
We took the scale parameter $d$ of Eq.~(\ref{eq2}) as $d=$ 3 fm$^{-1}$
and used a Gauss-Legendre quadrature with $N=$ 20 points.

\section{Results and discussion}
\label{secV}
\begin{figure*}[b]
\vspace*{-1cm}
\hspace{-2cm}
\resizebox{11.cm}{17.cm}{\includegraphics{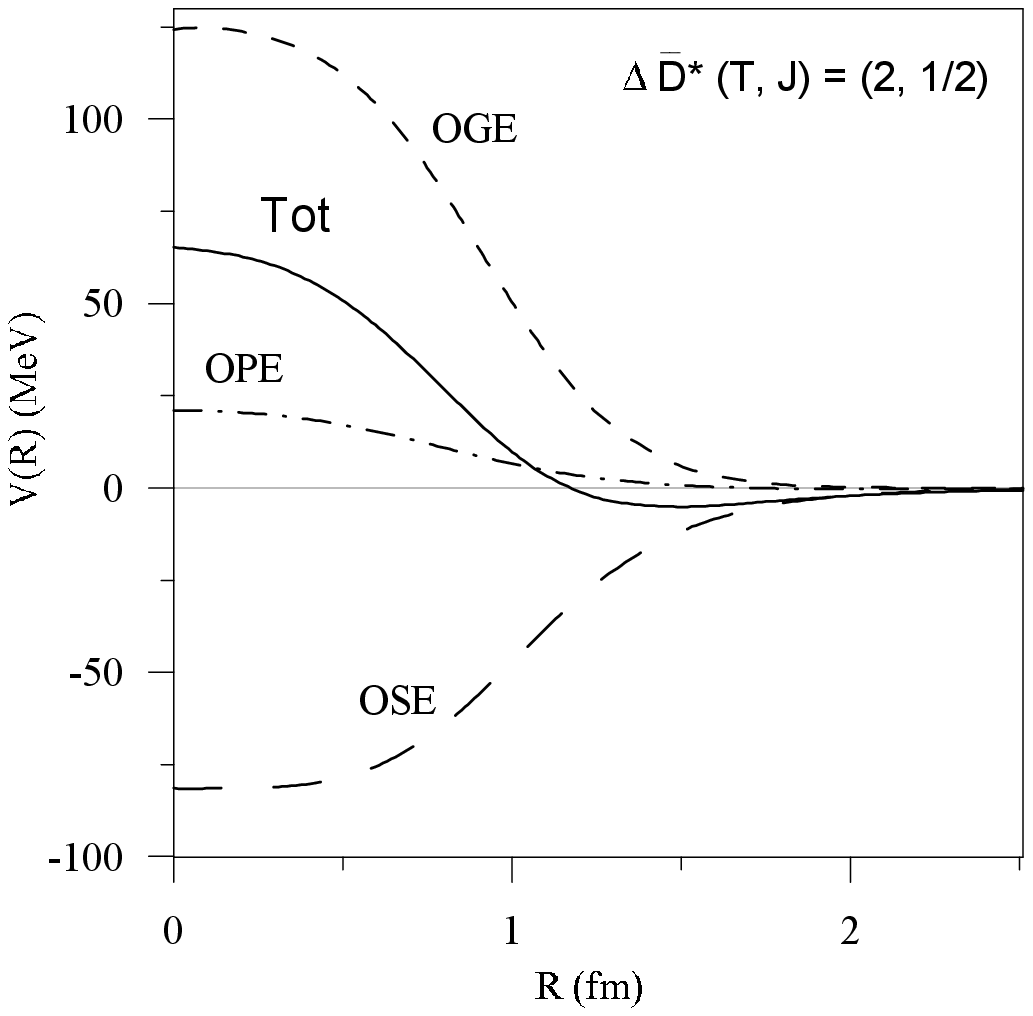}}\hspace{-4cm}
\resizebox{11.cm}{17.cm}{\includegraphics{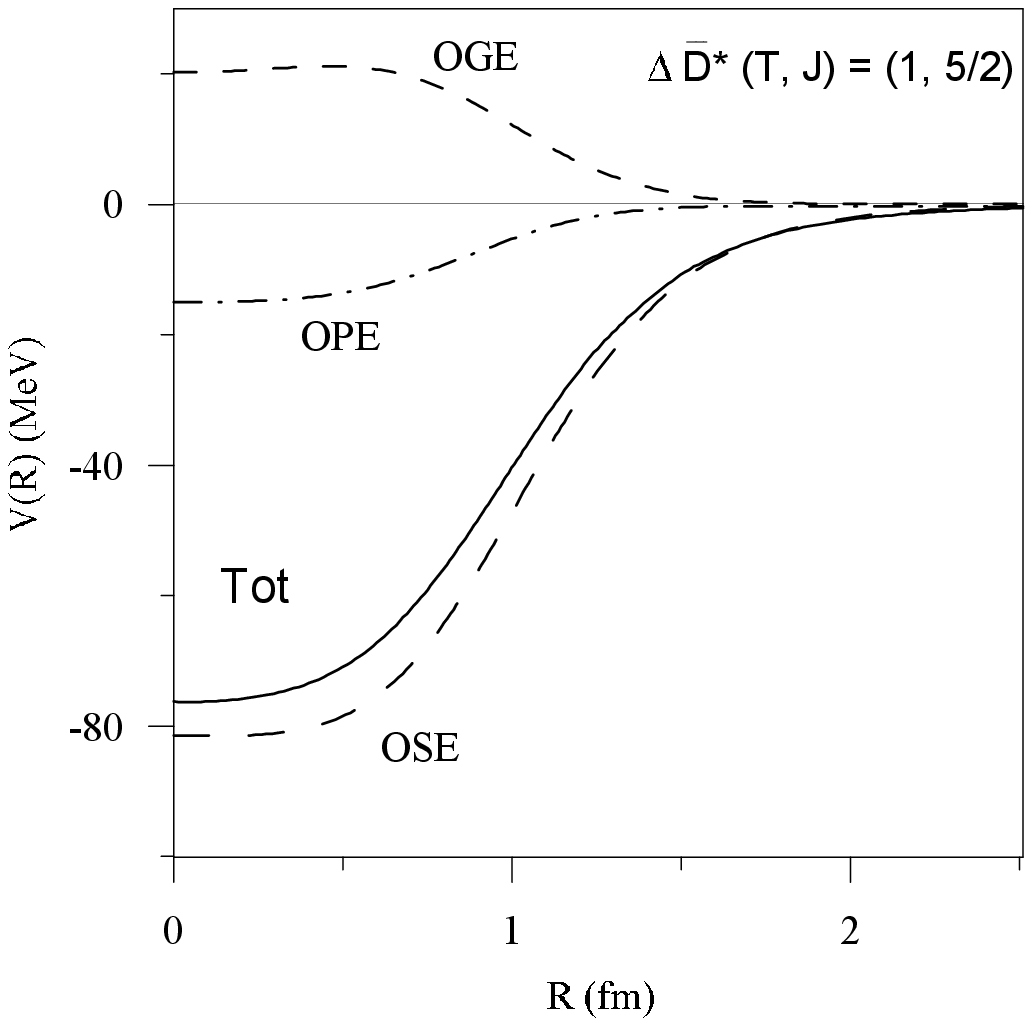}}
\vspace*{-10cm}
\caption{Contributions to the $\Delta \bar D^*$ interaction from the different pieces 
of the potential in two uncoupled channels: $(T,J)=(2,1/2)$ (left panel) and
$(T,J)=(1,5/2)$ (right panel).}
\label{fig3}
\end{figure*}
\begin{figure*}[t]
\vspace*{-1cm}
\hspace{-2cm}
\resizebox{11.cm}{17.cm}{\includegraphics{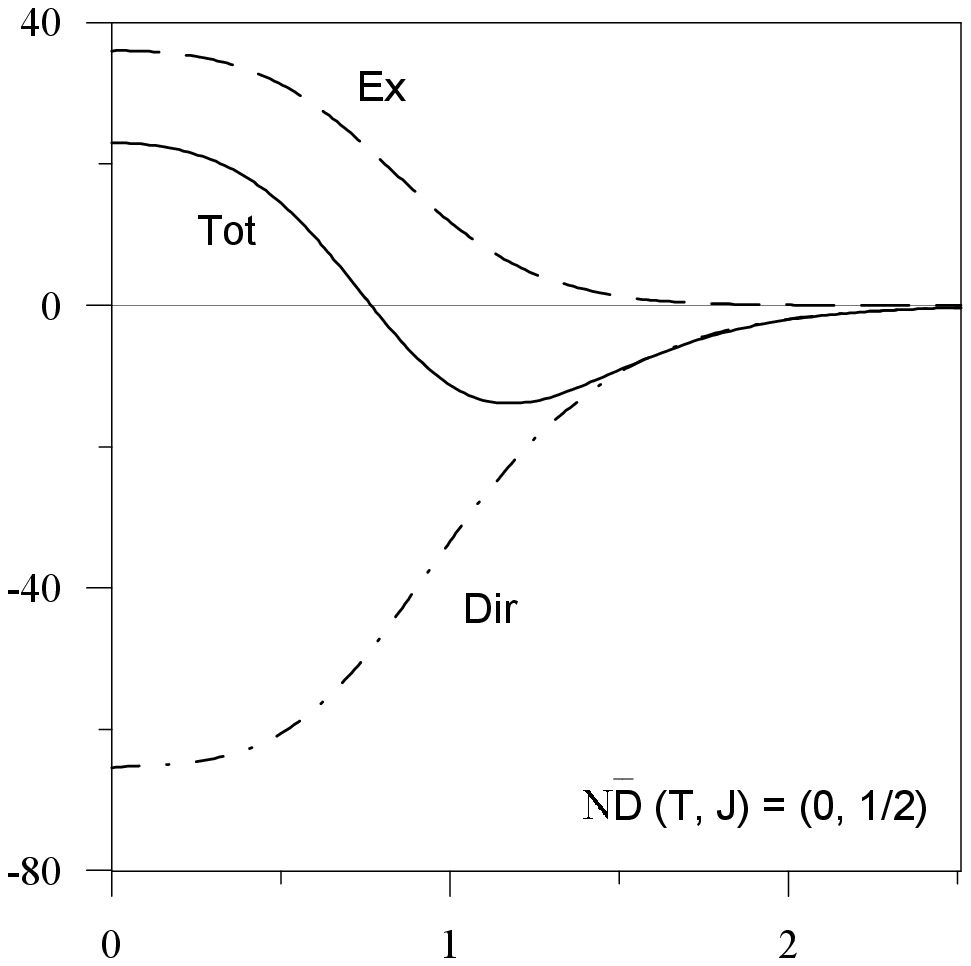}}\hspace{-4cm}
\resizebox{11.cm}{17.cm}{\includegraphics{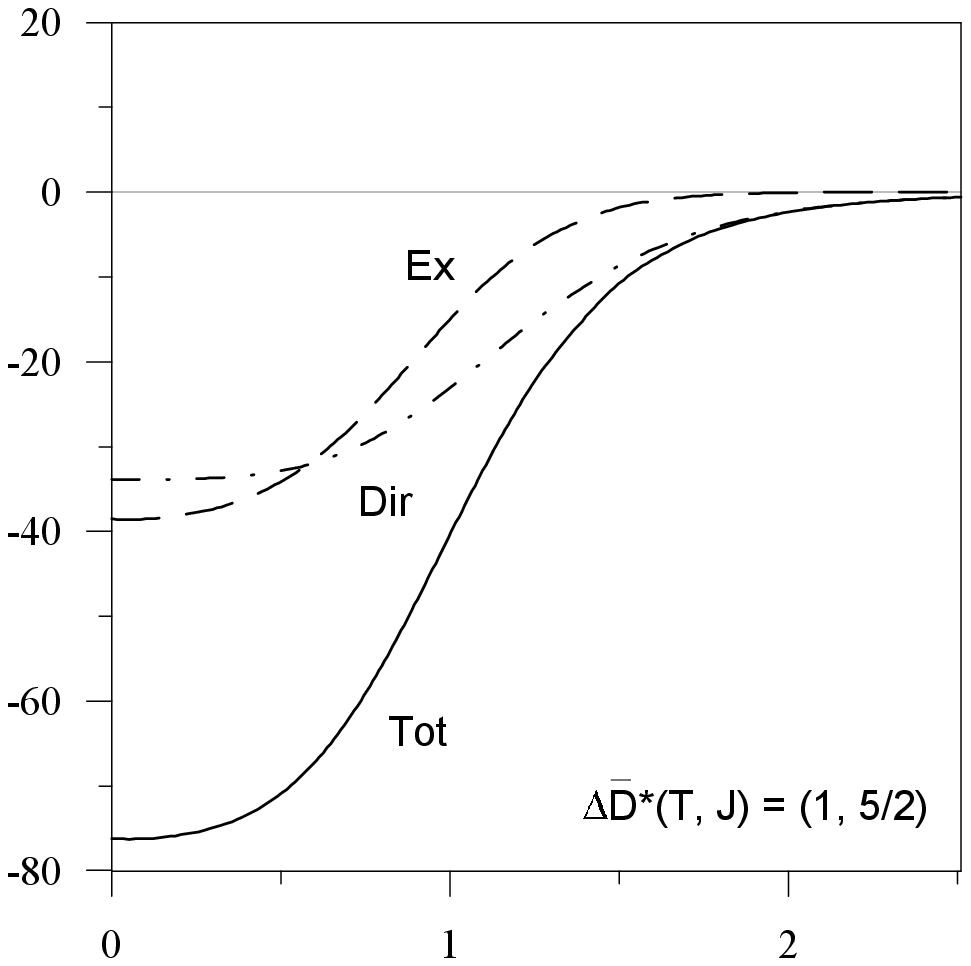}}
\vspace*{-10cm}
\caption{Direct and exchange contributions to the $N \bar D$ $(T,J)=(0,1/2)$ (left panel)
and $\Delta \bar D^*$ $(T,J)=(1,5/2)$ (right panel) potentials. The dashed-dotted line represents
the contribution of the direct terms, the dashed line stands for the exchange terms, and the solid 
line indicates the total potential.}
\label{fig4}
\end{figure*}

In Figs.~\ref{fig3} and~\ref{fig4} we show some representative potentials of the baryon-meson system
under study. In Fig.~\ref{fig3} we have depicted for two channels the contribution of the 
several pieces of the interacting potential: OGE, OSE and OPE. In Fig.~\ref{fig4} we have 
separated for two different channels the contribution of the different diagrams of Fig.~\ref{fig2} 
in two groups: direct terms, those shown in the first line that do not contain quark exchanges, 
and exchange terms, diagrams in the second and third lines of Fig.~\ref{fig2} where quark-exchange 
contributions appear. 

Regarding Fig.~\ref{fig3}, there are general trends
that have already been noticed for other two-hadron systems and we can briefly summarize.
For very-long distances ($R>$ 4 fm) the interaction comes determined by
the OPE potential, since it corresponds to the longest-range piece.
The OPE is also responsible altogether with the OSE for the
long-range part behavior (1.5 fm $<R<$ 4 fm), due to the combined effect of
shorter range and a bigger strength for the OSE as compared to the OPE.
The OSE gives the dominant contribution in the intermediate range (0.8 fm $<R<$ 1.5
fm), determining the attractive character of the potential in this region.
The short-range ($R<$ 0.8 fm) potential is either repulsive or attractive depending on the balance
between the OGE and OPE. In general, one can say that the OGE is mainly repulsive while the OSE 
is mostly attractive, see the two examples in Fig.~\ref{fig3}. Thus, in most cases the 
character of the OPE at short range determines the final character of 
the total potential.

It is also interesting to analyze the interaction in terms of the different diagrams 
plotted in Fig.~\ref{fig2}, as it has been done in Fig.~\ref{fig4}. The dynamical 
effect of quark antisymmetrization can be 
estimated by comparing the total potential with the one arising from the diagrams
in the first line of Fig.~\ref{fig2}, which are the only ones that do not include 
quark exchanges. Let us note
however that Pauli correlations are still present
through the norm, in the denominator of Eq.~(\ref{Poten2}). To eliminate the
whole effect of quark antisymmetrization one should eliminate quark-Pauli
correlations from the norm as well. By proceeding in this way one gets a
genuine baryonic potential, that we call direct potential. The comparison of
the total and direct potentials reflects the quark antisymmetrization effect
beyond the one-hadron structure.
In Fig.~\ref{fig4} we have separated  the contribution of the direct and exchange terms
for two different partial waves. As can be seen the
direct potential is always attractive, due both to the contribution of the scalar
exchange interaction and the absence of the one-gluon exchange (contributing only 
through quark-exchange diagrams). However, the character of the exchange part, containing quark-exchange
diagrams that would therefore be absent in a pure hadronic description, depends on the
sign of the color-spin-isospin coefficients. The sign of the dominant quark-exchange diagram,
$V_{34}P_{34}$, is crucial for determining whether the exchange contributions are
attractive or repulsive. Thus, quark-exchange diagrams give repulsion for the
$N \bar D$ $(T,J)=(0,1/2)$ channel (see left panel of Fig.~\ref{fig4}), but attraction for the 
$\Delta \bar D^*$ $(T,J)=(1,5/2)$ (see right panel of Fig.~\ref{fig4}), determining the final character of the
interacting potential.
Thus, dynamical quark-exchange effects play a relevant role in the
$N\bar D$ interaction with observable consequences as we will discuss below, being
responsible for the repulsive character of the $N\bar D$ interaction at short distances in
some partial waves.
\begin{figure*}[t]
\vspace*{-1cm}
\hspace{-2cm}
\resizebox{11.cm}{17.cm}{\includegraphics{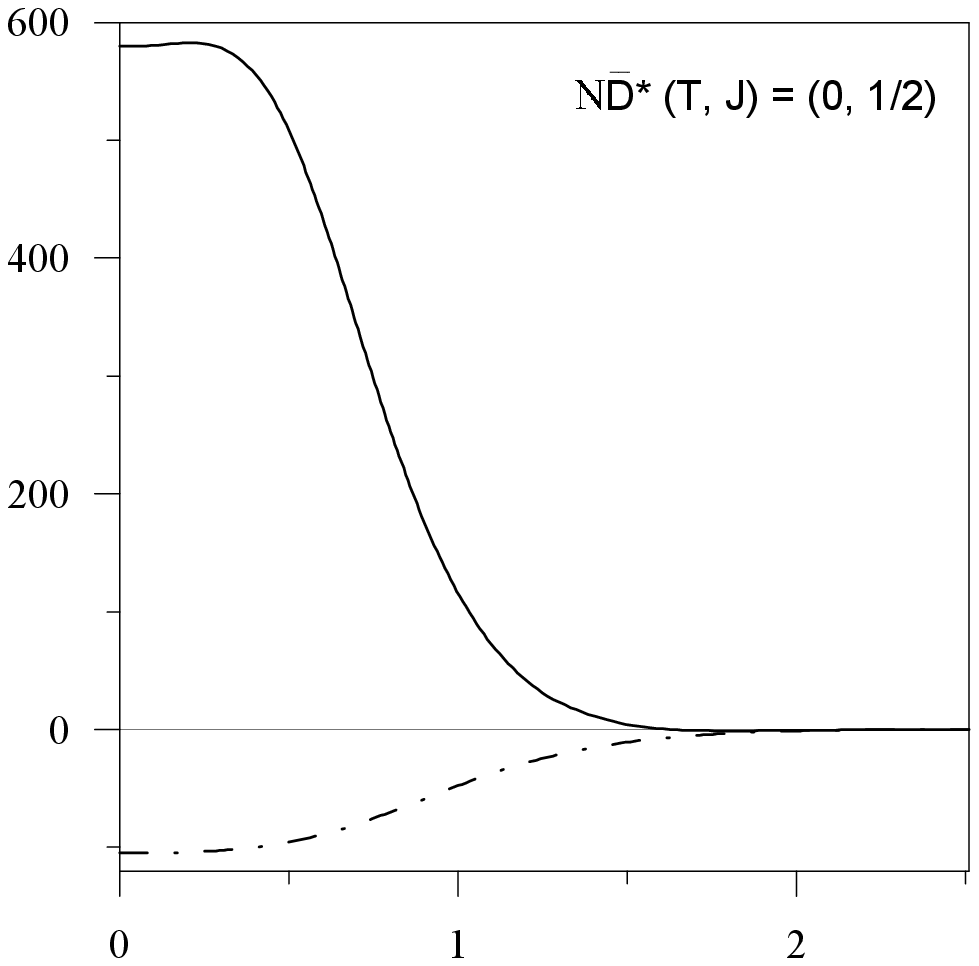}}\hspace{-4cm}
\resizebox{11.cm}{17.cm}{\includegraphics{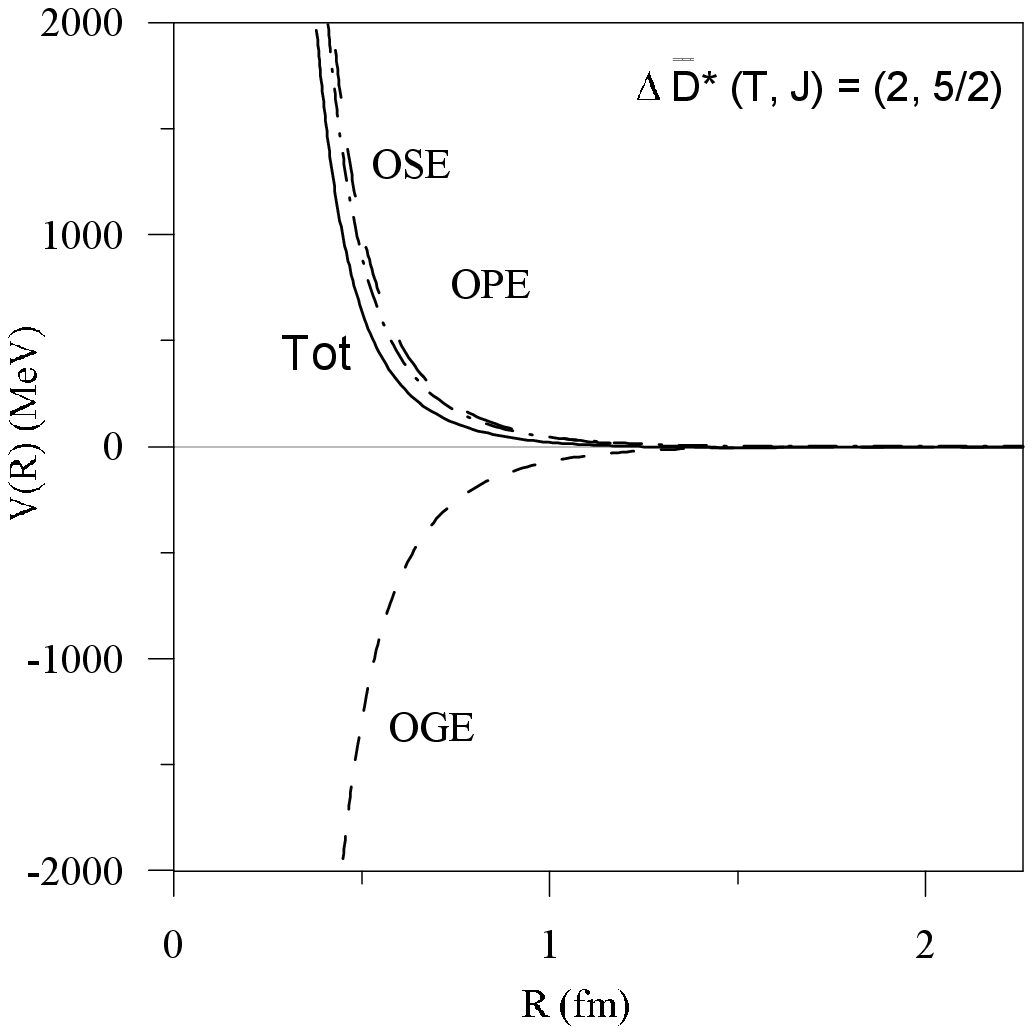}}
\vspace*{-10cm}
\caption{Interacting potential in two different channels showing Pauli effects: $N \bar D^*$ 
$(T,J)=(0,1/2)$ (left panel) and  $\Delta \bar D^*$ $(T,J)=(2,5/2)$ (right panel). 
See text for details.}
\label{fig5}
\end{figure*}
\begin{table}[b]
\caption{Character of the interaction in the different baryon-meson $(T,J)$ channels.}
\label{tab4}
\begin{tabular}{cccc}
\hline
\hline
& $T=0$  &  $T=1$ & $T=2$ \\
\hline
$J=1/2$ &  Repulsive           &  Repulsive         &  Weakly repulsive      \\
$J=3/2$ &  Weakly attractive   &  Weakly repulsive  &  Attractive            \\
$J=5/2$ &                      &  Attractive        &  Strongly repulsive    \\ \hline \hline
\end{tabular}
\end{table}

The potential becomes strongly repulsive in those cases where we have Pauli blocked, $C(S,T)=1$,
or Pauli suppressed channels, $C(S,T)$ close to one. In the left panel of Fig.~\ref{fig5} we have drawn 
the $N \bar D^*$ $(T,J)=(0,1/2)$ interaction. As seen in Table~\ref{tab1} the value
of $C(S,T)=2/3$ is close to one, suppressing the overlapping of the wave function
at short distances, see Fig~\ref{fig1}. This gives rise to a strong repulsion at 
short range. The dashed-dotted line in this figure represents the direct potential, the one
obtained at hadronic level neither considering exchange diagrams in the norm kernel
nor in the interacting potential. As can be seen, the effect of quark-exchanges is 
rather important in what we have called Pauli suppressed partial waves. Hadronic 
interactions would therefore not be able to
account for the consequences of Pauli effects as it was already noticed
in a comparative study of the $N\Delta$ interaction by means of
hadronic or quark-based models~\cite{Val94}.
In the right panel of Fig.~\ref{fig5} we have drawn the $\Delta \bar D^*$ $(T,J)=(2,5/2)$ interaction.
In this case $C(S,T)$ is exactly one, forbidding the overlapping of the two 
hadrons for $R=0$. All contributions are very strong at short distances due to the
behavior of the norm kernel and the total interaction becomes strongly repulsive. These
forbidden states would have to be eliminated by hand in the RGM treatment of the two-hadron
systems~\cite{Fuj07}. The existence of such a strong repulsion has also been observed in the
$N\Delta$ system and may be concluded from the $N\Delta$ phase shift behavior derived
from the $\pi d$ elastic scattering data~\cite{Fer90}.
\begin{figure*}[t]
\vspace*{-1cm}
\hspace{-2cm}
\resizebox{11.cm}{17.cm}{\includegraphics{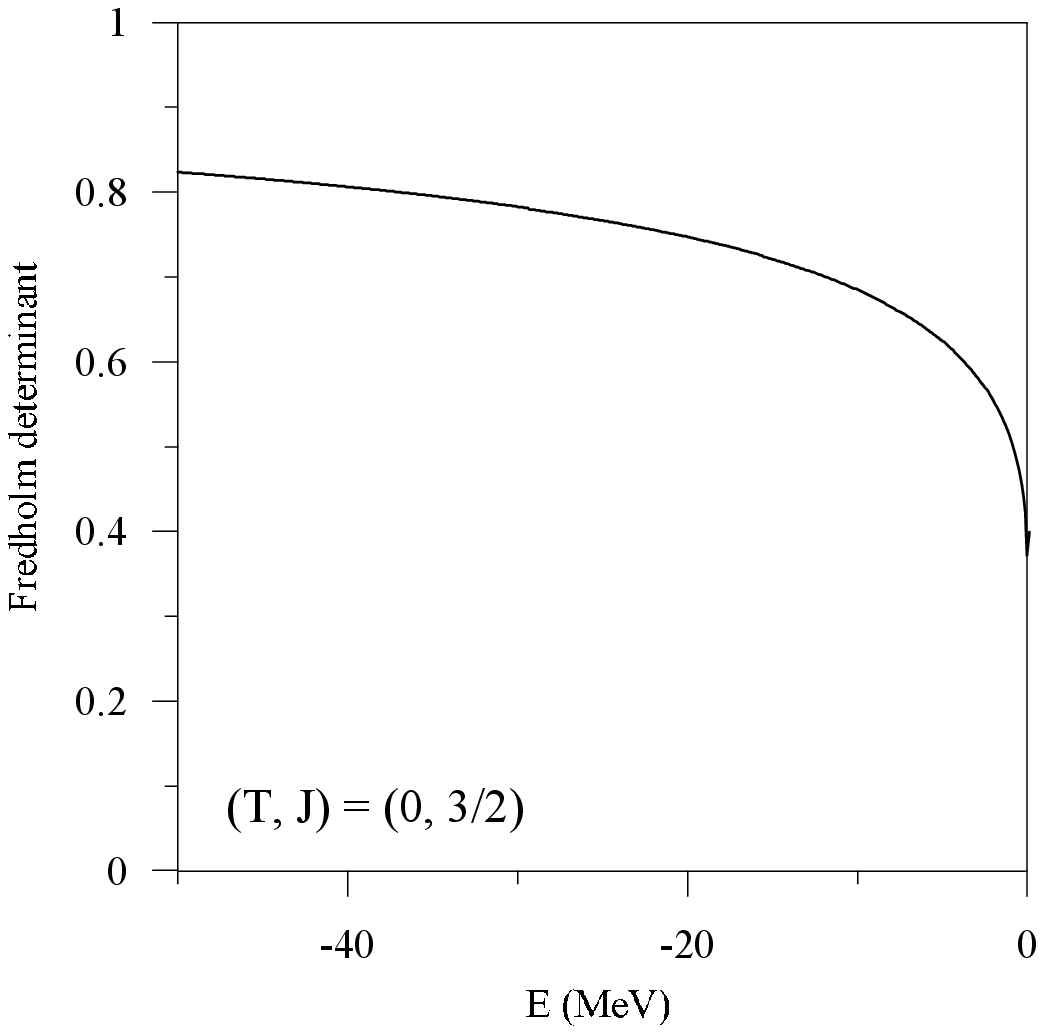}}\hspace{-4cm}
\resizebox{11.cm}{17.cm}{\includegraphics{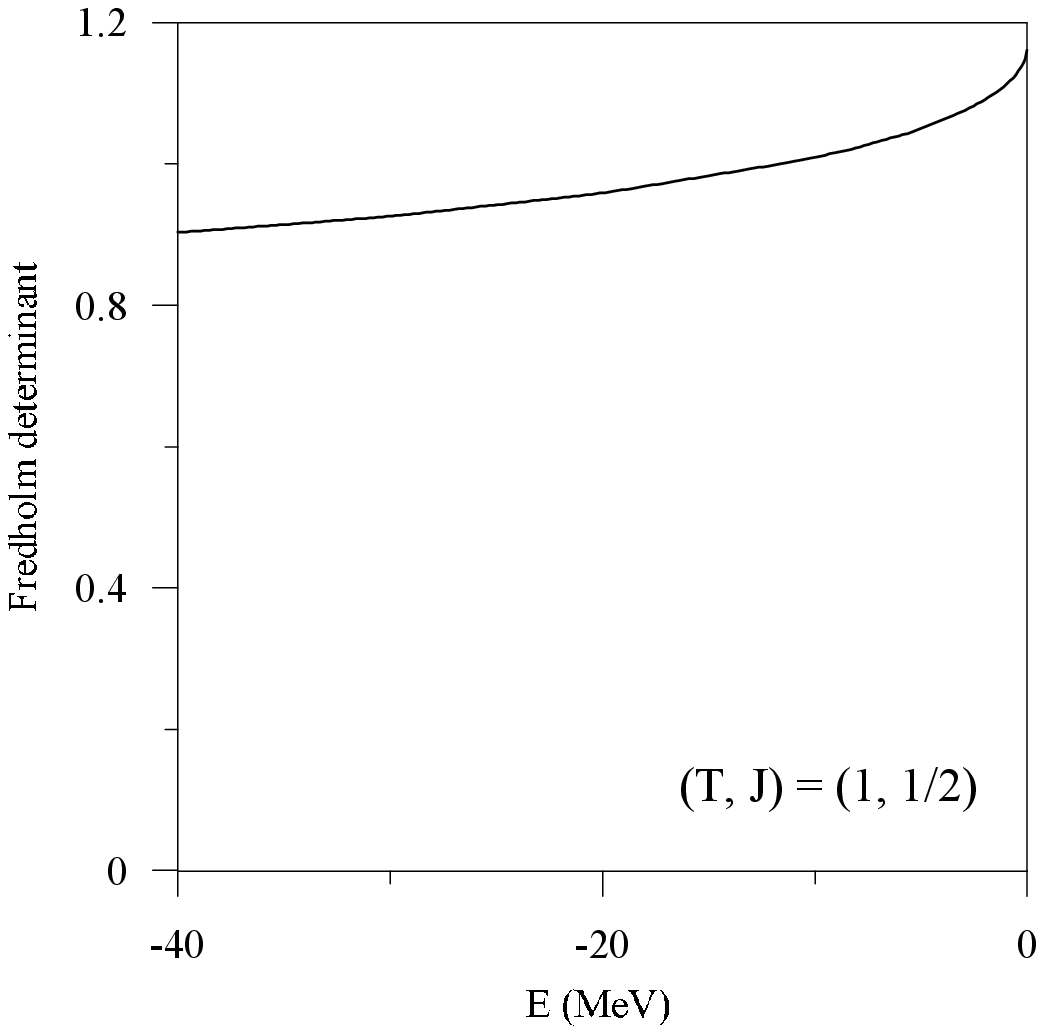}}
\vspace*{-10cm}
\caption{$(T,J)=(0,3/2)$ (left panel) and $(T,J)=(1,1/2)$ (right panel) Fredholm determinant.}
\label{fig6}
\end{figure*}
\begin{figure}[b]
\vspace*{-1cm}
\hspace*{-1cm}\mbox{\epsfxsize=160mm\epsffile{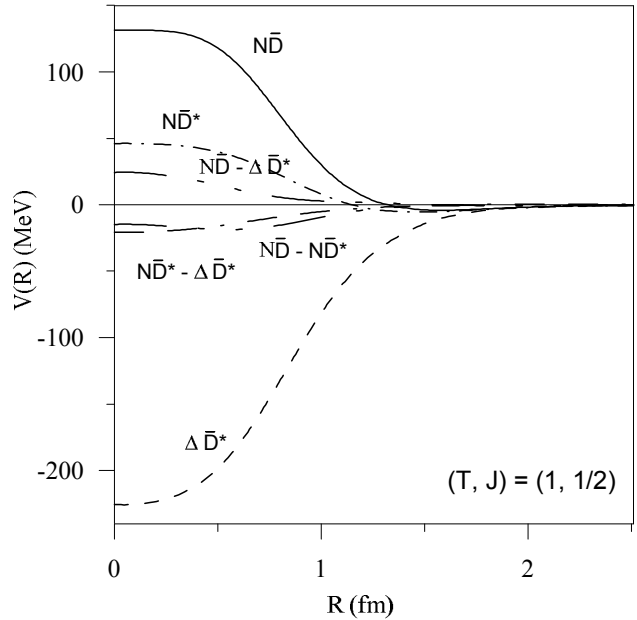}}
\vspace*{-13.5cm}
\caption{Baryon-meson potentials contributing to the $(T,J)=(1,1/2)$
channel.}
\label{fig7}
\end{figure}
\begin{figure*}[t]
\vspace*{-1cm}
\hspace{-2cm}
\resizebox{11.cm}{17.cm}{\includegraphics{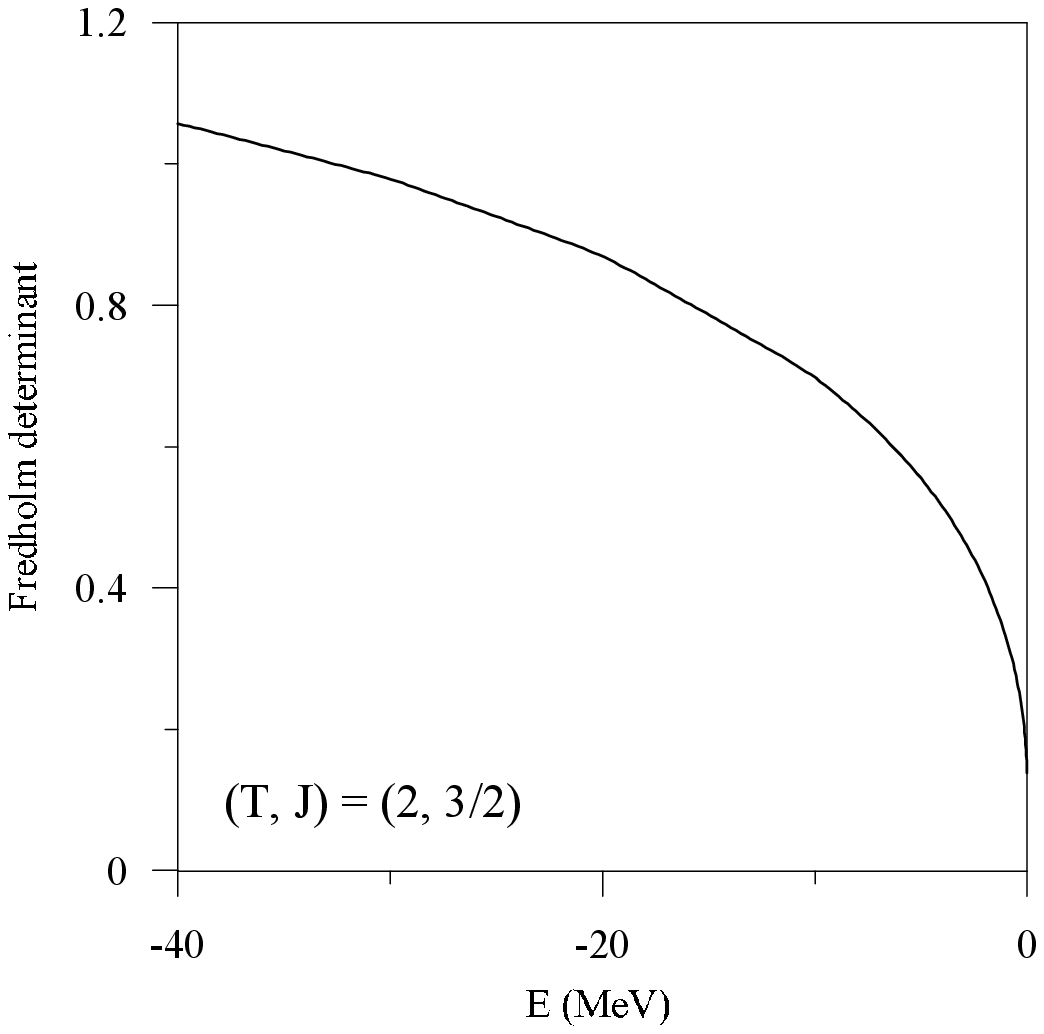}}\hspace{-4cm}
\resizebox{11.cm}{17.cm}{\includegraphics{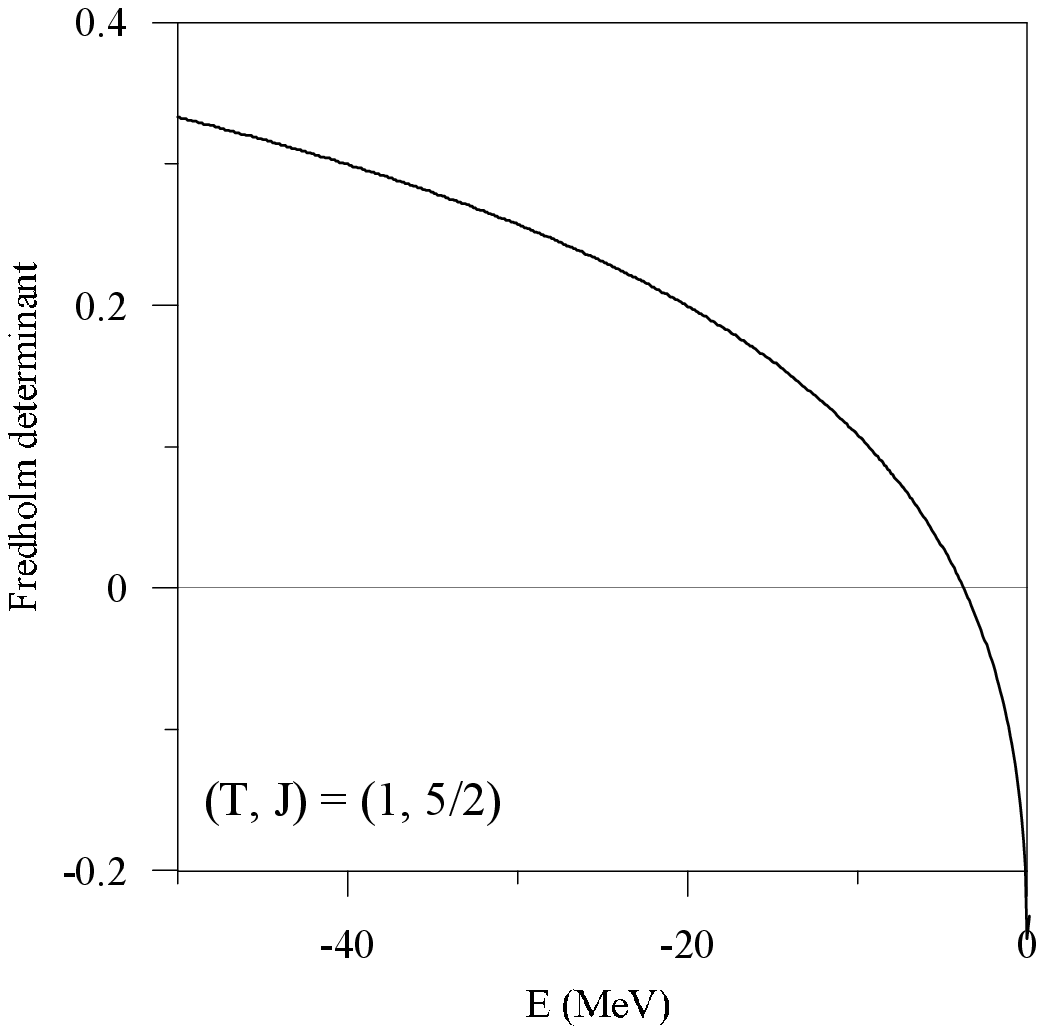}}
\vspace*{-10cm}
\caption{$(T,J)=(2,3/2)$ (left panel) and $(T,J)=(1,5/2)$ (right panel) Fredholm determinant.}
\label{fig8}
\end{figure*}

Using the interactions described above, we have solved the coupled channel
problem of the baryon-meson systems made of a baryon, $N$ or $\Delta$, and a
meson, $\bar D$ or $\bar D^*$ as explained in Sec.~\ref{secIV}. The existence of 
bound states or resonances will generate exotic states with charm $-1$ 
that could be identified in future experiments of the $\bar {\rm P}$ANDA Collaboration at 
the FAIR facility~\cite{Pan09}. In Table~\ref{tab4} we summarize 
the character of the interaction in the different $(T,J)$ channels.
It can be observed how all $(T,J)$ channels containing Pauli blocked or
Pauli suppressed states are repulsive: $(2,5/2)$, $(0,1/2)$, and $(1,1/2)$. 
Thus, the Pauli principle at the level
of quarks plays an important role in the dynamics of the $N\bar D$ system.
\begin{figure}[b]
\vspace*{-1cm}
\hspace*{-1cm}\mbox{\epsfxsize=160mm\epsffile{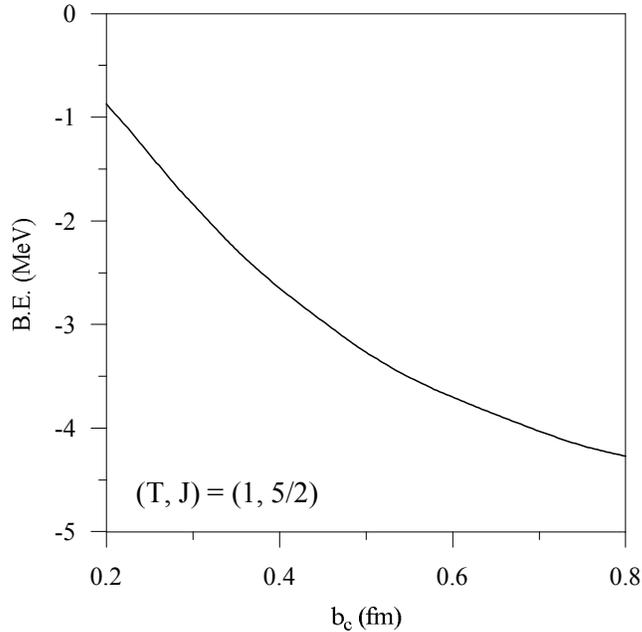}}
\vspace*{-13.5cm}
\caption{Binding energy, in MeV, of the $(T,J)=(1,5/2)$ channel as
a function of the harmonic oscillator parameter used for the charm quark.}
\label{fig9}
\end{figure}

In Figs.~\ref{fig6} and~\ref{fig8} we have plotted the Fredholm determinant for some
representative channels. Fig.~\ref{fig6} shows the weakly attractive $(T,J)=(0,3/2)$ and
the weakly repulsive $(T,J)=(1,1/2)$ Fredholm determinants. In Fig.~\ref{fig7} we have
plotted all potentials contributing to the $(T,J)=(1,1/2)$ channel that, as
seen in the right panel of Fig.~\ref{fig6}, is weakly repulsive. The interaction in the lightest 
two-hadron systems contributing to this channel, $N\bar D$
and $N\bar D^*$, has a repulsive character and there is a weak coupling to the 
attractive but heavier two hadron system $\Delta \bar D^*$, giving an overall repulsive
channel. Fig.~\ref{fig8} shows the Fredholm determinant for the two attractive
channels of the $N\bar D$ system: $(T,J)=(2,3/2)$ and $(T,J)=(1,5/2)$. The $(T,J)=(1,5/2)$ is the most attractive one, 
showing a bound state with a binding energy of 3.87 MeV. This channel corresponds to a
unique physical system, $\Delta \bar D^*$. This situation is rather similar to the one
we found in the $\Delta \Delta$ system~\cite{Val01}, predicting an $S$ wave resonance
with maximum spin. Experimental evidence for such a resonance
was already reported in the $NN$ scattering data, where the resonance appeared in the $^3D_3$
partial wave, thus as a $D$ wave in the $NN$ system~\cite{Arn00}.
This prediction has been recently used as a possible explanation of the
measured cross section of the 
double-pionic fusion of nuclear systems through the so-called
{\it ABC} effect~\cite{Bas09}. The formation of an intermediate $\Delta\Delta$ resonance with the isospin, spin, 
parity and mass found in Ref.~\cite{Val01} ($(T)J^P=(0)3^+$ and $M=$2.37 GeV) allowed to describe
the cross section of the double-pionic fusion
reaction $pn \to d \pi^0\pi^0$. In the present case, the bound state in the 
$(T,J)=(1,5/2)$ $\Delta \bar D^*$ channel would appear in the scattering of
$\bar D$ mesons on nucleons as a $D$ wave resonance, what could in principle be
measured at $\bar {\rm P}$ANDA. Such a state will have quantum numbers $(T)J^P=(1)5/2^-$.
Just to make sure that our conclusion does not depend on our choice of the parameter
of the charm quark wave function, that does not come out from the solution of the meson spectra,
we have calculated the binding energy as a function of $\rm{b_c}$. The result
appears in Fig.~\ref{fig9}, showing a bound system for any reasonable choice of 
such a parameter, giving thus confidence to our prediction.
\begin{figure}[t]
\vspace*{-1cm}
\hspace*{-1cm}\mbox{\epsfxsize=160mm\epsffile{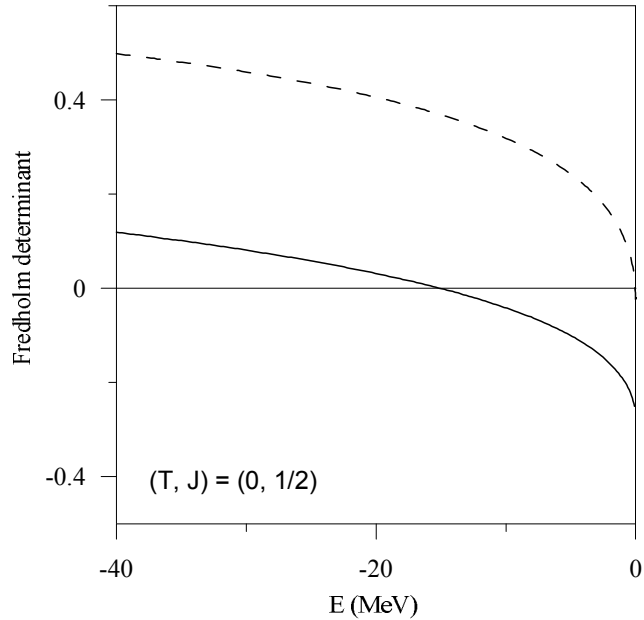}}
\vspace*{-13.5cm}
\caption{$(T,J)=(0,1/2)$ Fredholm determinant obtained with the direct potentials.
The dashed line stands for the single channel problem $N\bar D$ and the solid line
to the couple channel problem using also the $N\bar D^*$ state.}
\label{fig10}
\end{figure}

In Refs.~\cite{Yas09,Yam11} the $N\bar D$ system has been analyzed by means of a hadronic model
using Lagrangians satisfying heavy quark symmetry and chiral symmetry. They arrive to the
conclusion that the $(T)J^P=(0)1/2^-$ channel is the most attractive one presenting a bound
state of around 1.4 MeV. To emphasize the importance of quark-exchange effects we have repeated 
for this channel the
calculations explained in Sec.~\ref{secIV} using only the direct potentials, those  without quark-exchange effects. These
potentials, that would correspond to a purely baryonic interaction, 
are drawn as a dashed-dotted line in the left panels of Fig.~\ref{fig4}, the $(T,J)=(0,1/2)$ $N\bar D$
interaction, and Fig.~\ref{fig5}, the $(T,J)=(0,1/2)$ $N\bar D^*$ interaction. In both cases
the importance of quark-exchange effects can be seen. These interactions
would be the hadronic potentials of an effective theory without quark degrees of freedom.
The results obtained neglecting the contribution
of quark-exchange effects are shown in Fig.~\ref{fig10}.
As we can see in both cases, single channel or coupled-channel calculation, 
a bound state appears, in agreement with
the conclusions of Refs.~\cite{Yas09,Yam11}. Once again, this comparison makes evident
the great importance that quark-exchange effects may have in the system under study
and it also represents a sharp example of a system where the quark-exchange dynamics 
may have observable consequences. A future effort in the study of the $N\bar D$ system will provide
us with evidence to learn about the importance of quark-exchange dynamics.

Let us finally mention that there are recent estimations in the literature~\cite{Hai07} about
the $N\bar D$ cross section based on hybrid models considering meson-exchanges supplemented
with a short-range quark-gluon dynamics. The important conclusion of that work is that
the predicted $N \bar D$ cross sections are of the same order of magnitude as those
for the $NK$ system, but with average values of  20 mb, roughly a factor two larger
than for the latter system. Thus, the study of the $N\bar D$ interaction could be 
a plausible challenge for the $\bar P$ANDA Collaboration. 

\section{Summary}
\label{secVI}

Summarizing, we have studied the $N \bar D$ system at low energies by means of a chiral 
constituent quark model that describes the baryon and meson spectra as well as the $NN$ interaction.
Due to the presence of a heavy antiquark the interaction becomes rather simple and
parameter-free predictions can be obtained for the $N\bar D$ system. 
We have analyzed in detail the interaction in the different isospin-spin channels,
emphasizing characteristic features consequence of the contribution of quark-exchange dynamics.
Quark-Pauli effects generate a strong repulsion in some particular channels
due to lacking degrees of freedom to accommodate the light quarks.
Such effects have observable consequences generating repulsion in channels that
otherwise would be attractive in a hadronic description. We have traced back
our results to the previous analysis of the $N\Delta$ and $\Delta\Delta$ systems 
with peculiar predictions supported by the experimental data.
We have found only one bound state in the 
$\Delta \bar D^*$ $(T,S)=(1,5/2)$ system. This state, $(T)J^P=(1)5/2^-$, will show up 
in the scattering of $\bar D$ mesons on nucleons as a $D$ wave resonance. Such a resonance resembles
our findings in the $\Delta\Delta$ system that offered a plausible explanation to the cross section
of double-pionic fusion reactions through the so-called
{\it ABC} effect. The existence of this state is a sharp prediction of quark-exchange
dynamics because in a hadronic model the attraction appears in different channels.
Finally, it is important to emphasize that theoretical estimations indicate that 
the $N\bar D$ cross section may be attainable in the future facility FAIR and 
therefore the predicted resonance may be a challenge for the study of the 
$\bar {\rm P}$ANDA Collaboration. This objective may constitute a helpful tool
in discriminating among the different scenarios used to describe the dynamics of 
heavy hadron systems.

\acknowledgments
This work has been partially funded by the Spanish Ministerio de
Educaci\'on y Ciencia and EU FEDER under Contract No. FPA2010-21750,
and by the Spanish Consolider-Ingenio 2010 Program CPAN (CSD2007-00042).

\end{document}